\documentclass[10pt,english,twocolumn]{article}
\usepackage[T1]{fontenc}
\usepackage[latin9]{inputenc}
\usepackage{geometry}
\usepackage{babel}

\usepackage{color}
\PassOptionsToPackage{normalem}{ulem}
\usepackage{ulem}

\usepackage{authblk}
\usepackage{float}
\usepackage{graphicx}

\makeatletter
\usepackage{lmodern}
\date{}

\makeatother

\begin{document}

\title{Simple cellular automata to mimic foraging ants submitted to abduction}

\author{F. Tejera\thanks{Correspondence to: ftejera@fisica.uh.cu} }
\author{E. Altshuler}

\affil{Physics Faculty, "Henri Poincar\`{e}" Group of Complex Systems, University of Havana, Cuba}
\affil{To download the published version of this paper, go to:\\
http://rcf.fisica.uh.cu/index.php/es/2015-07-15-15-25-47/volumen-32-numero-1-2015}
\maketitle

Modeling complex physical phenomena through computer simulations has become a useful tool for understanding the world around us. Even the simplest realistic models base on the general laws of physics, aimed to solving large systems of partial differentials or integro-differentials equations, and represent exceptional numerical problems. One of the techniques used to address these problems are called Cellular Automata (CA). In brief, they are simulations based on simple rules in which the space, time, and the possible states of the system are discrete \cite{VonNeumann1966, Wolfram1994,Bandini2010}.

Biological systems, especially moving groups of animals, provide many key examples in collective phenomena that can be modeled by CA \cite{Ermentrout93,cole1996,John2004}. Most of these collective phenomena show two well defined levels of organization which are the individual level of the organism and the overall level of a group. Most experiments reveal the features of the global level, but determining which individual scale interactions are involved in creating, for example, self-organized patterns, represents a tough challenge \cite{Altshuler2005,Noda2006,John09,Nicolis2013}.

Ants belong to the group of animals that organize without centralized control: in fact, they are a paradigm of collective behavior. Living in society involves both cost and benefit. For example, they pay the cost related to the high density of individuals living together in a common space, but they benefit from the access to the information handled by their nestmates \cite{Jackson2004a}. The high density of ants in a foraging trial results in a decrease in their average velocity  and therefore in a decrease in  the flow of food returning to the nest. This is due among other things, to the large number of frontal head-encounters that occur between ants moving to and from the food source. 
These encounters, on the other hand, are expected to contribute to the information exchange between individuals. Since ants are often subjected to numerous threats during foraging (as could be from a predator or adverse weather events such as rain \cite{Farji-brener2010,Prabhakar2012,Halley2001}), it is reasonable to believe that head-head encounter are crucial to exchange danger information. In the present work we investigate, using CA models, the emergent patterns resulting from different hypothesis of ant-ant exchange of danger information when foragers are abducted.

Our simulations represent an ant colony foraging for food though a trail including ants going from the nest to the foraging area (out-bound ants) and ants returning to the nest from the foraging area (nest-bound ants). During a given time interval ants are abducted with a certain probability at a point in the trail, simulating the presence of a predator. The non-abducted ants just change the direction of motion and move towards the nest carrying the information of danger. In our model, we assume different hypothesis on their ability to share this information with their nestmates moving to the foraging area. To study how the colony responds to abduction, we analyze what happens to the flow of ants leaving the nest looking for food. We expected that the exchange of danger information between ants would result in a decrease of the number of out-bound ants as a protection mechanism.

As we have mentioned, CA are discretized models, in which the continuous space is replaced by an array of cells. In our case, the size of each cell is such that only an ant can occupy a cell at each instant of time: it corresponds to 2 cm in reality. Ants positions, therefore, may only be changed in discrete steps, that will be integer multiples of the cell size.

\begin{figure}[h]
\includegraphics[scale=0.53]{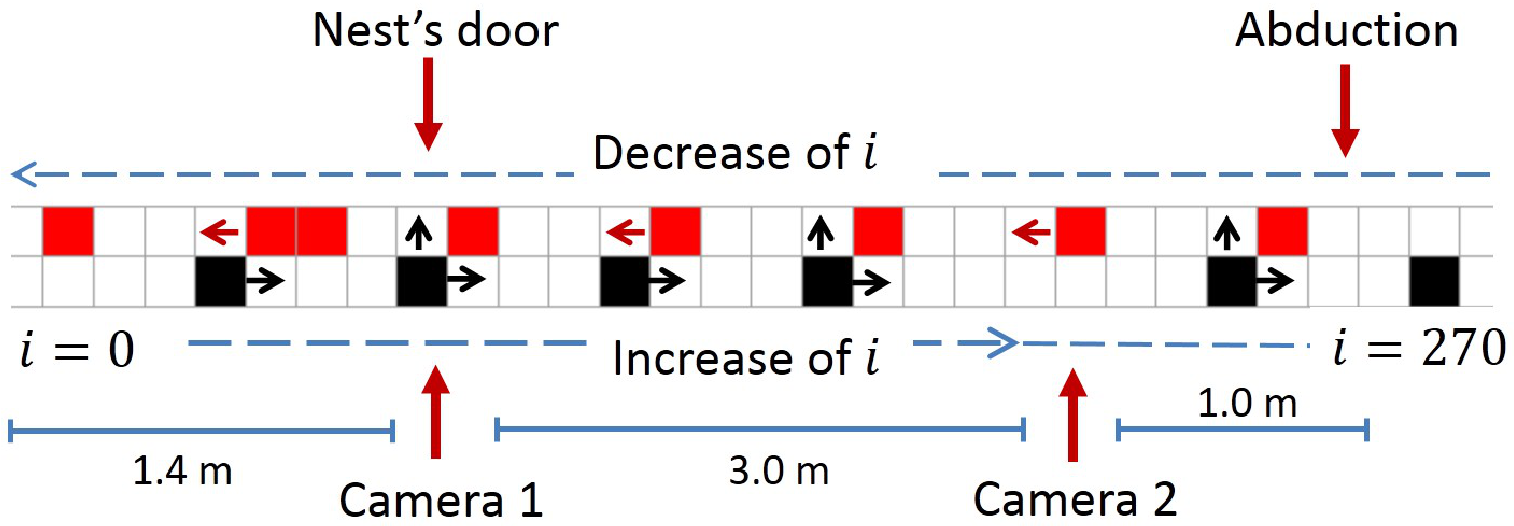}
\caption{Schematic diagram of the CA foraging trail, including the corresponding dimensions in a realistic experiment. The trail is observed using two cameras. Out-bound-ants and nest-bound ants are represented by black squares and red squares, respectivelly.}\label{fig: schematic diagram}
\end{figure}

Time is also increased in discrete amounts. In our model, the duration of each step is 1 second. All virtual ants move with constant velocity equal to 2 cm/s. Therefore, the sequence of successive states in our CA, is like a sequence of photographs taken of the entire system. 

\begin{figure*}
\begin{center}
\includegraphics[scale=0.4]{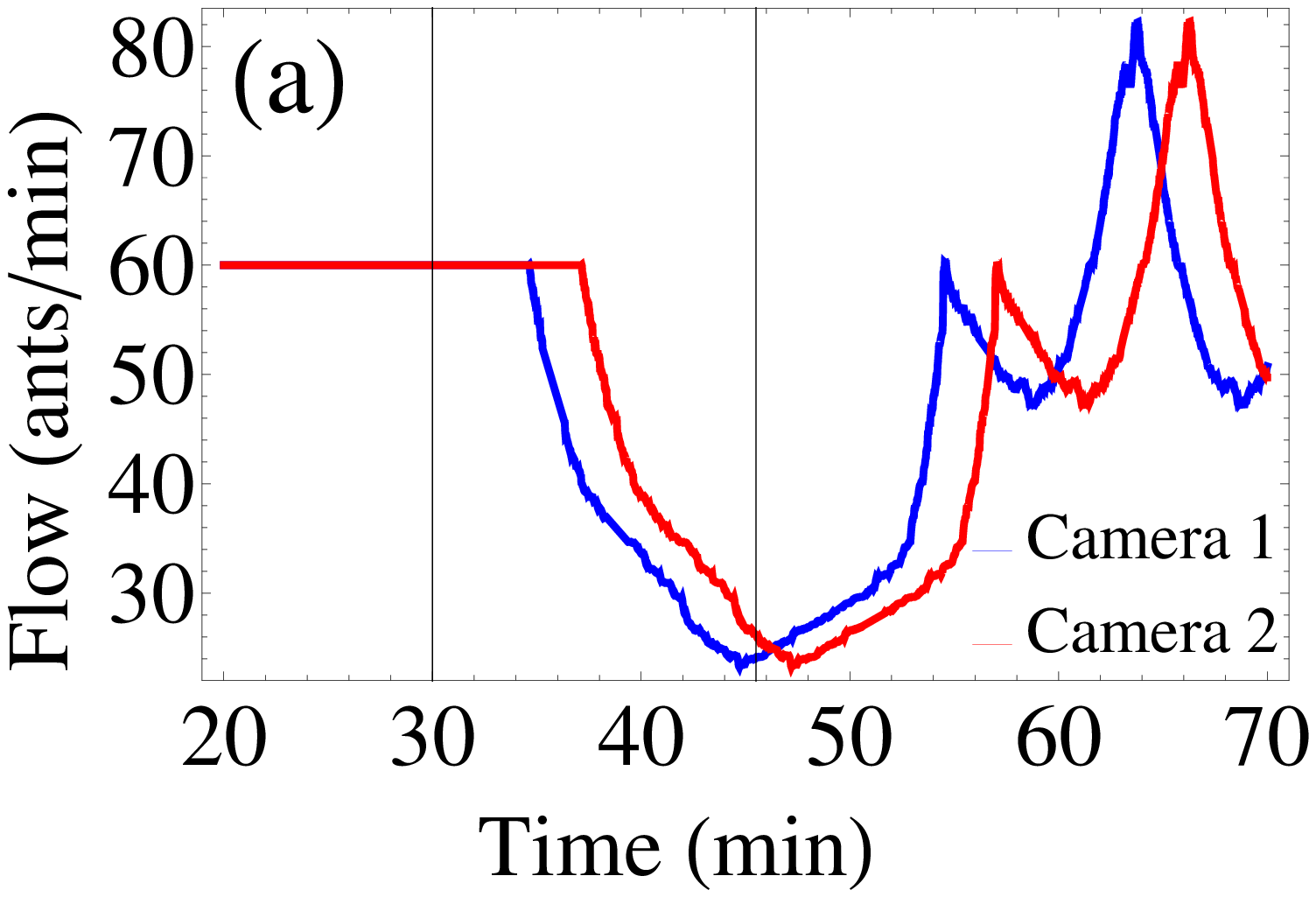}\includegraphics[scale=0.4]{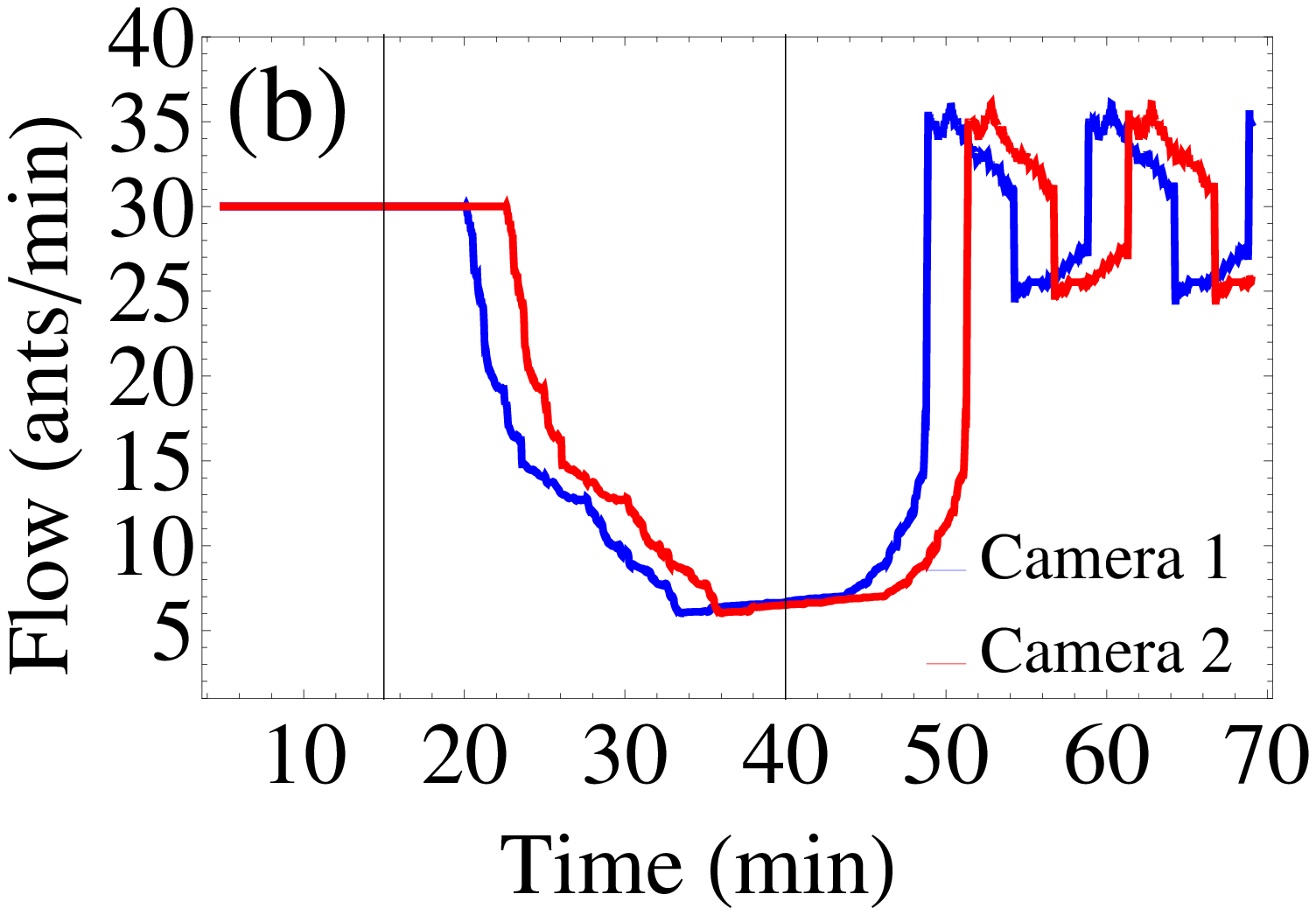}\includegraphics[scale=0.29]{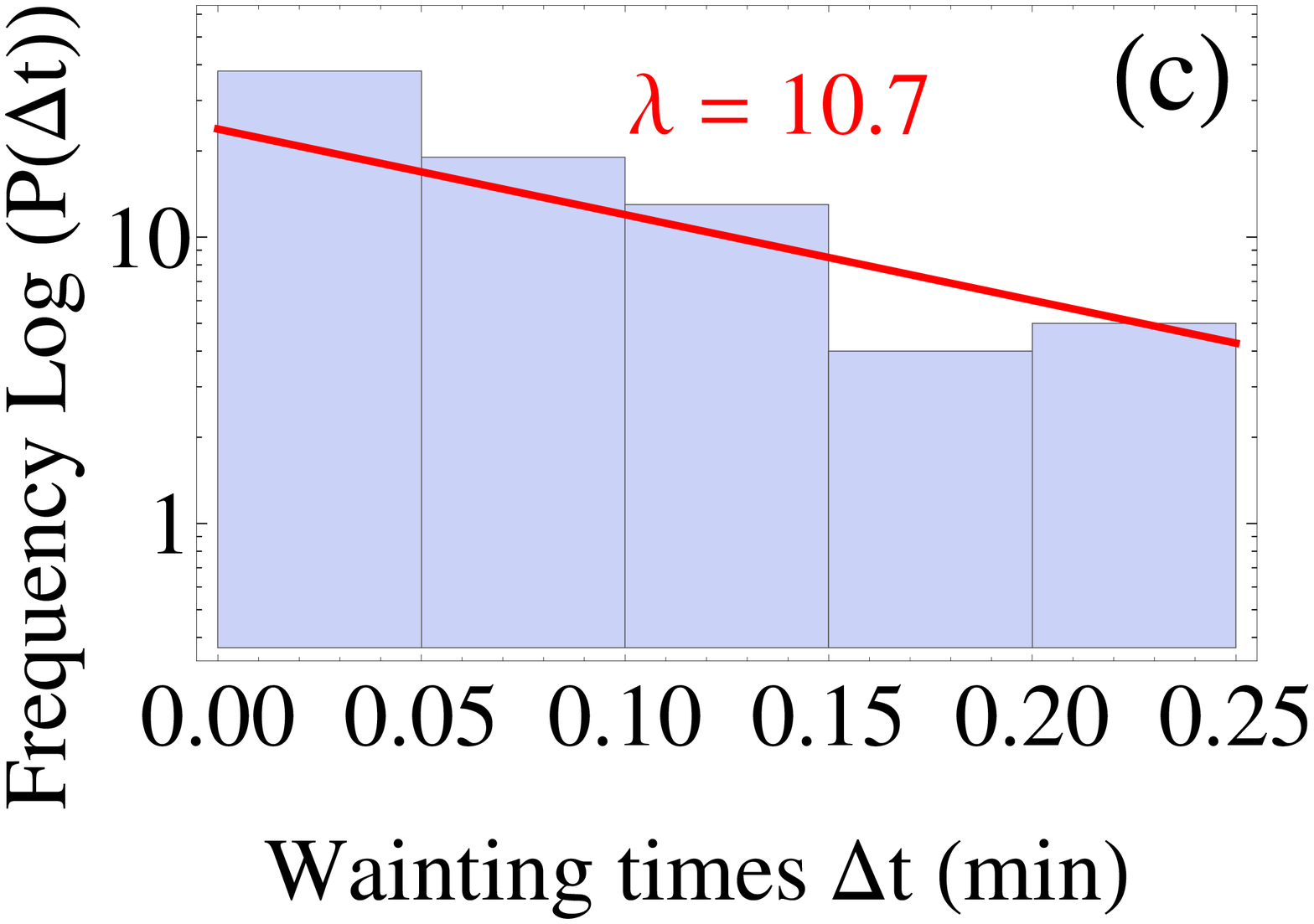}

\includegraphics[scale=0.4]{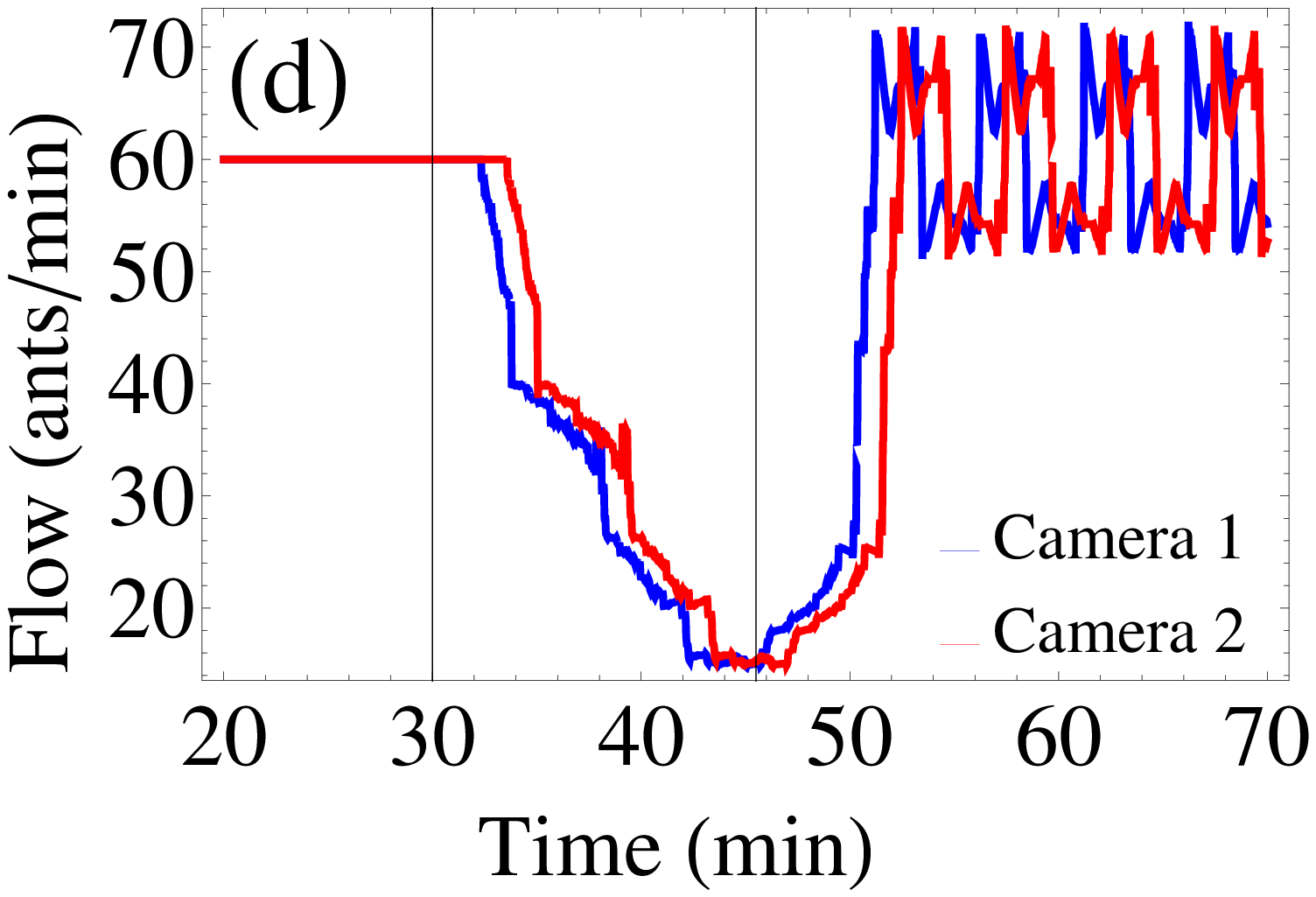}\includegraphics[scale=0.4]{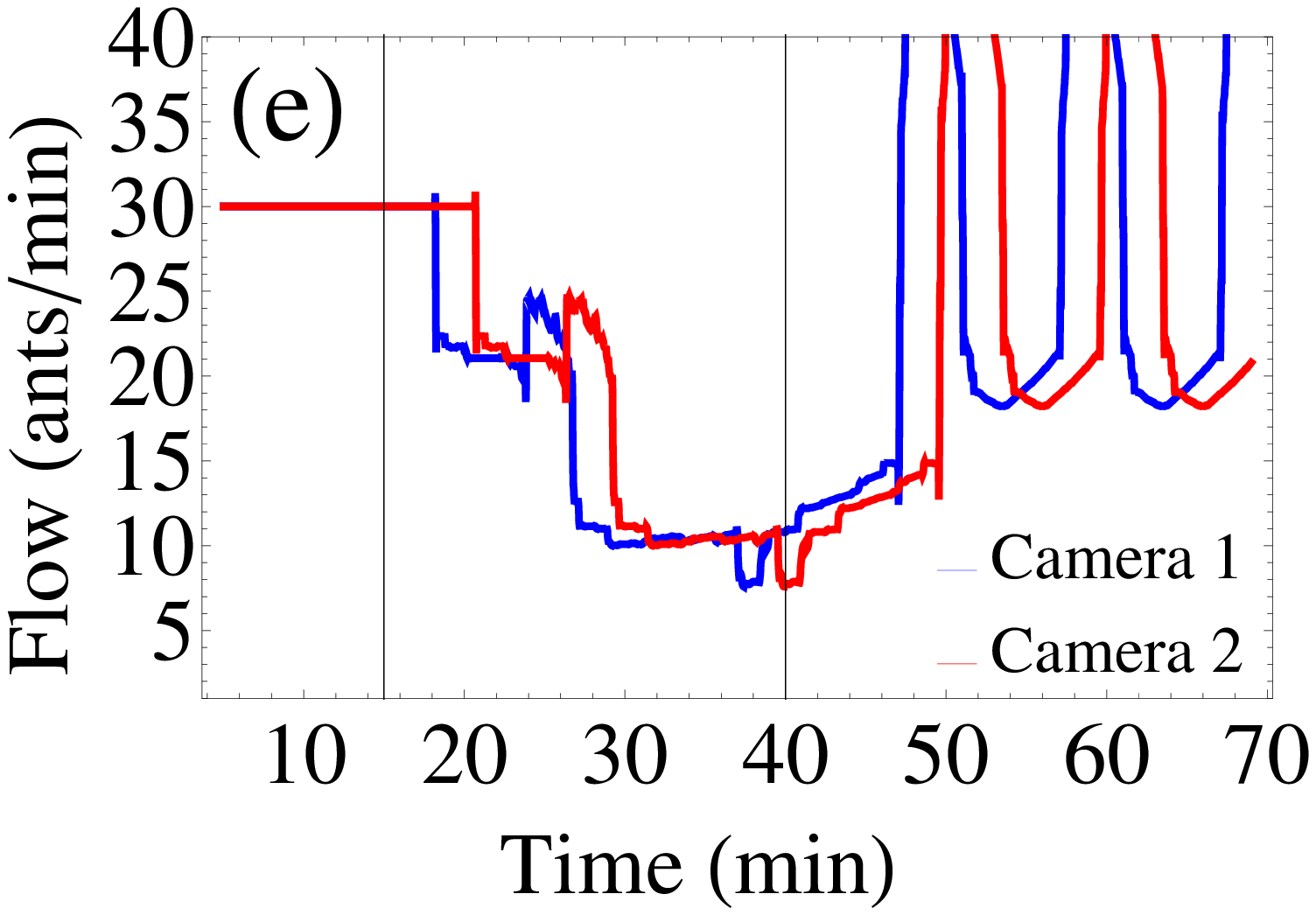}\includegraphics[scale=0.28]{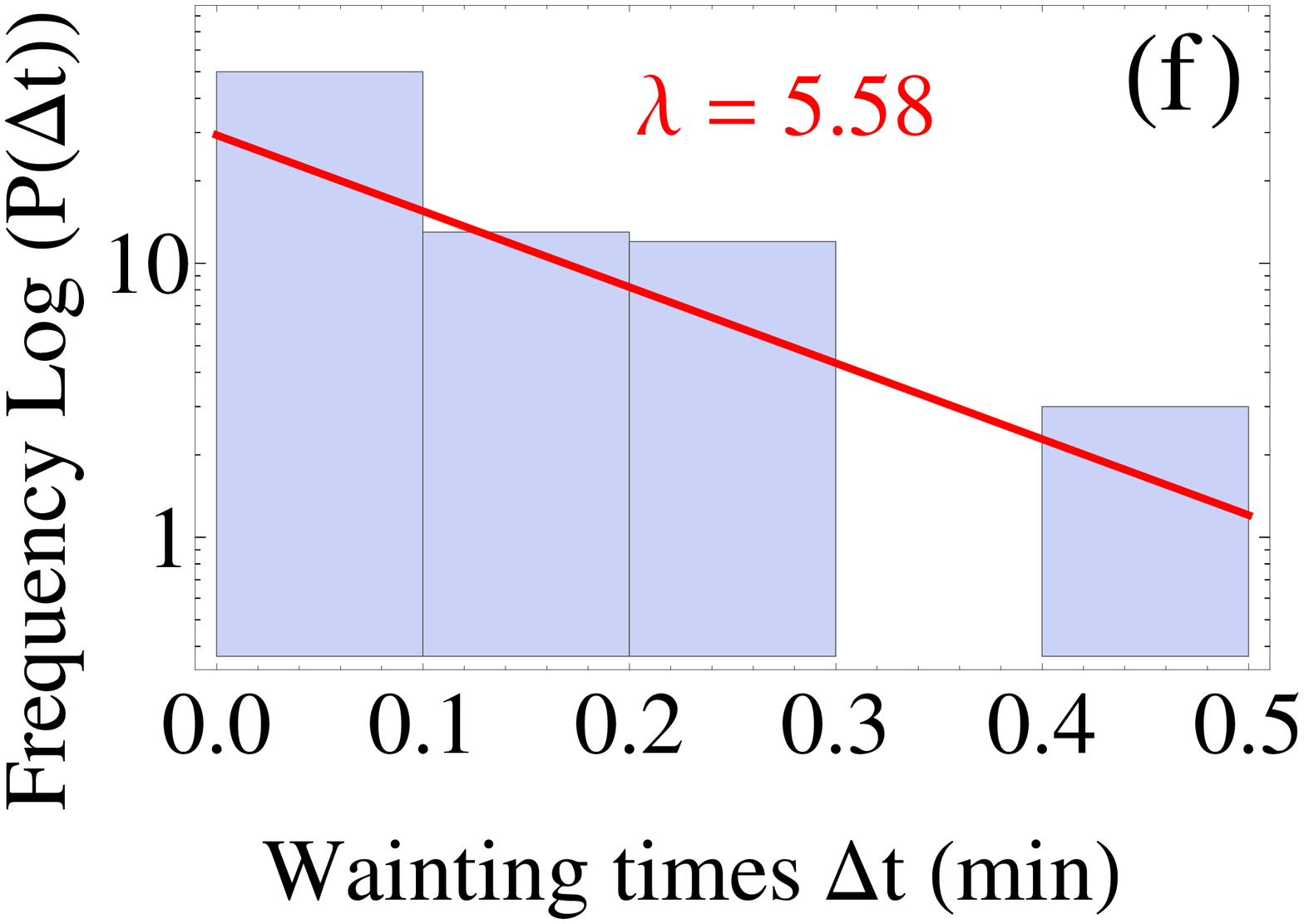}

\caption{Flows of out-bound ants passing by both cameras and waiting times distributions, simulated by CA. (a) and (d) Flows of out-bound ants passing by both cameras with 15 min of abduction, corresponding to $n_{max}p_{\alpha}=0$ and $n_{max}p_{\alpha}=0.8$, respectively. (b) and (e) Flows of out-bound ants passing by both cameras with 25 min of abduction, corresponding to $n_{max}p_{\alpha}=0$ and $n_{max}p_{\alpha}=0.8$, respectively. (c) and (f) Waiting times distributions for simulations with 25 min of abduction, corresponding to $n_{max}p_{\alpha}=0$ and $n_{max}p_{\alpha}=0.8$, respectively.}\label{fig: flows_activity}
\end{center}
\end{figure*}

The total duration of the simulations is approximately 1 hour, partitioned among before, during and after the kidnapping of ants. We studied two cases resembling actual experiment conditions. In the first the flow of ants is 60 ants/min and individual are kidnapped for 15 minutes (Experiment 1). In the second case the flow is 30 ants/min and ants are kidnapped for 25 minutes (Experiment 2). Ants can move in two directions, toward the food source or back to the nest as shown in Fig \ref{fig: schematic diagram}. During the time of abduction ants that reach the danger zone can be randomly removed from the line with a probability $P_{ab}=0.6$. So, for each ant that reaches the area of abduction, a random number (uniformly distributed) between 0 and 1 is generated and compared with the probability of being kidnapped. If:
\begin{equation}
\label{eq probabilidad de secuestro}
Random (0,1)>P_{ab},
\end{equation}
the ant is kidnapped. Otherwise the ant just reverses its directions of motion and returns to the nest along the top row shown in Fig. \ref{fig: schematic diagram}.
As the returning ants reach the nest, they move inside it for some extra distance, change direction, and go out again towards the abduction area (button row in Fig. \ref{fig: schematic diagram}). Each abducted ant is accumulated along the top row at the right of the abduction zone, to resemble the laden ants coming from the foraging area that cannot pass through the abduction area. When the abduction period ends, the ants accumulated at the right go back to the nest, thus contributing to the recovery of the flow.

Each line of the simulation contains 270 cells, equivalent to a total of 5.4 meters, divided into three parts between the nest and the area of abduction: over 1.4 m corresponding to the estimate of the length of ants move from the interior of the nest to its door (where camera 1 is located); 3 m corresponding to the distance between camera 1 and camera 2, and 1 m corresponding to the distance between camera 2 and abduction zone (see Fig. \ref{fig: schematic diagram}). Table \ref{tab paarmetros del CA} summarizes parameters' value used in the simulations.

\begin{table}
\caption{Paramters used in the simulations.\label{tab paarmetros del CA}}
\begin{center}
\begin{tabular}{|c|c|c|} 

\hline 
 Parameter & Experiment 1 & Experiment 2\tabularnewline
\hline 
Abduction time & $15\mbox{ min}$ & $25\mbox{ min}$\tabularnewline
\hline 
Average ant velocity & $v_{h}=2\mbox{cm/s}$ & $v_{h}=2\mbox{ cm/s}$\tabularnewline
\hline 
Linear density of ants & $0.5\mbox{ ants/cm}$ & $0.25\mbox{ ants/cm}$\tabularnewline
\hline 
Total number of ants & 600 & 300\tabularnewline
\hline 
Total trail length & $5.4\mbox{ m}$ & $5.4\mbox{ m}$\tabularnewline
\hline 
\end{tabular}

\end{center}
\end{table}

All ants returning from the area of abduction during the kidnapping's time, may carry danger information and may transmit it to the out-bound ants. Since there is no experimental evidence on danger information exchange outside the nest, we assume that the ants heading back to the nest from the abduction area, transmit the danger information only inside the nest. As consequence, out-bound ants are induced to perform "U" turns and move back inside the nest if enough danger information is received. Let us assume then that ants share hazard information only within the 1.40 m inside the nest. The status of each ant can be represented as follows:
\begin{eqnarray}
a_{n}(i,t+1) & \rightarrow & F(a_{n}(i,t))\label{eq: hormigas que regresan}\\
a_{o}(i,t+1) & \rightarrow & F(a_{o}(i,t),f(n,p_{\alpha}),a_{n}(i,t)).\label{eq:hormigas que salen}
\end{eqnarray}

Where $a_{n}(i,t)$ and $a_{o}(i,t)$ are the states of occupancy of the position \emph{i} at time \emph{t}, by nest-bound and out-bound ants, respectively ($a_{n}$ and $a_{o}$ only take the values 0 or 1). \emph{F} is the "motion function", that we describe as follows. In (\ref{eq: hormigas que regresan}), if position $i$ (in the upper lane of Fig. \ref{fig: schematic diagram}) is not occupied at time \emph{t}, the nest-bound ant at the position $i+1$ moves into $i$ at time steps $t+1$, but it does not move if the cell $i$ is occupied. In (\ref{eq:hormigas que salen}), an analogous rule holds along the lower lane, but there is an important difference. If an out-bound ant at position \emph{i} in the lower lane, coincides with a nest-bound ant in the upper lane, it will increase its "danger information" by adding 1 to the parameter $n$. This value multiplied by the "panic factor" $p_{\alpha}$ represents the "survival instinct", and increases as ants meet her companions who survived abduction. If it satisfied that, 
\begin{equation} 
Random (0.1) <np _ {\alpha},
\label{eq: comparison}
\end{equation} 
ant \emph{i} in the lower lane will try to jump to the upper lane (equivalent to perform a "U-turn" and return to the nest). This process is included in the function \emph{f}, in (\ref{eq:hormigas que salen}). But that is only possible if position \emph{i} at time \emph{t} is not occupied in the upper lane. This is why $a_{n}(i,t)$ is present in (\ref{eq:hormigas que salen}).

The panic factor $ p_{\alpha} $, characterizes the intensity of danger information communicated by returning ants. By tuning this factor we can regulate the proportions between the needs to find food and to protect individuals from danger. At the same time $p_{\alpha}$
is taken in such a way that $0\leq n_{max}p_{\alpha}\leq 1$, to avoid that all ants decide to return after a threshold number of ants encounters. Here, $n_{max}$ represents the maximum number of contacts that an ant may experiences inside the nest.

We first analyze the extreme case in which the ants do not share information, and therefore never induce U-turns. This is true for $ p_{\alpha} = 0$.

Fig. \ref{fig: flows_activity} (a) and (b), show the values for the flow of out-bound ants seen by the two cameras spaced 3 meters apart, corresponding to both kidnapping intervals. We observe that  before the abduction period the system is in a stationary state. During abduction the ants flow decreases due to abduction, and afterwards the flow recovers.

These general features are also observed in the case of $n_{max}p_{\alpha}=0.8$, see Fig. \ref{fig: flows_activity} (d) and (e). The oscillations found in the flow after abduction period, are due to "avalanche effects": if once an ant is informed of danger and decides to return, it communicates the information to their nestmates, leading to a multiplicative process.

\begin{figure}
\includegraphics[scale=0.43]{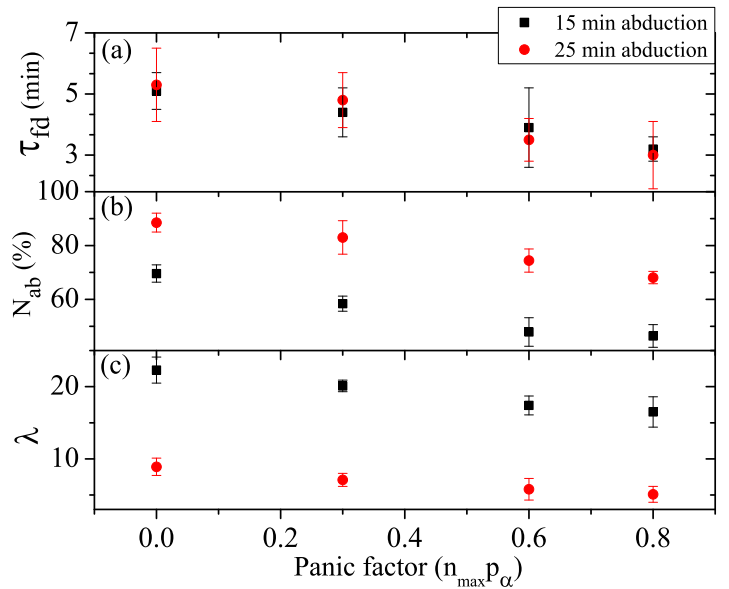}
\caption{Simulation outputs as a function of the panic parameter. (a) Time delay between the beginning of abduction and start of the flow decrease, $\tau_{fd}$. (b) Percentage of abducted ants, $N_{ab}$. (c) Exponent of the distributions of waiting times during abduction, $\lambda$.} \label{fig: juntos}
\end{figure}

However, we note the following important differences. The time when the flow begins to decrease once kidnapping begins ($\tau_{fd}$), decreases with increasing  $ n_{max}p_ {\alpha}$, as shown in Fig. \ref{fig: juntos} (a). This suggest that increasing the intensity $p_{\alpha}$, ants in the nest learn faster about danger and begin to make U-turns. In fact, this results in a decrease of the number of ants being kidnapping in the abduction area, see Fig. \ref{fig: juntos} (b).

The simulations show that even although the panic factor is set to the maximum value there is necessary some time (around three minutes) for the danger information to reach the nest. Also, the model as proposed in (\ref{eq: hormigas que regresan}) and (\ref{eq:hormigas que salen}), does not permit that the colony suppresses completely the flow even for the maximum panic factor (Fig. \ref{fig: flows_activity} (d) and (e)), i.e  there will always be  some ``kamikaze" scouts willing to forage. This model gives flexibility to the colony to ``prioritize" foraging. We expect that these two parameters, $\tau_{fd}$ and $ n_{max}p_ {\alpha}$, allow direct comparison with experimental data.

Another important quantity that characterizes the traffic on the line of ants is the temporal spacing between ants (or waiting times).
This parameter is defined as the
difference of passage time between an ant ({\it i}) and its nearest nestmate ($i+1$): 
$\Delta t=t_{i+1}-t_{i}$ and gives us an idea of the ant spatial distribution along the row, during abduction period. The distributions of waiting times for the out-bound ants for the simulations with 25 min of abduction, can be described by a Poisson process, i.e. by exponential distributions $P(t) = e^{-\lambda t}$, as show in Fig \ref{fig: flows_activity} (e) and (f). This means that waiting times are mutually independent: there are no correlations between two successive waiting times. During the abduction we reduced the density of ants on the line, so we see longer waiting times for the out-bound ants in the case of maximum panic factor, which is reflected as a smaller slope of the distribution (plotted in a log-linear graph, \ref{fig: flows_activity} (e) and (f)). If we increase the panic factor or the time of abduction, there is a  decrease in the value of the exponent $\lambda$, corresponding to longer waiting times. Fig. \ref{fig: juntos} (c) shows the decrease of $\lambda$ as $n_{max}p_{\alpha}$ increases. This is also a fingerprint of the mechanism against danger.

Our model shows that the more information is exchanged between members of the colony the bigger will be the response to external stimuli, meaning more protection of their members. But, at the same time shows that the colony does not stop foraging.

The parameters we have used in the simulations have been estimated for actual experiments under natural conditions. So, it would be easy to test our hypothesis in real experiments. Preliminary experimental results (which will be published elsewhere) suggest that the behavior of foraging ants of species \emph{Atta insularis} is best described using the hypothesis $p_\alpha=0$. So, surprisingly enough, real ants do not share danger information in abduction experiment.

\bibliographystyle{abbrvnat}

\begin{thebibliography}{10}

\bibitem{VonNeumann1966}
{Von Neumann}, J.,
\newblock {\em {Theory of Self-Reproducing Automata}},
\newblock University Press, 1966.

\bibitem{Wolfram1994}
Wolfram, S.,
\newblock {\em {Cellular Automata and Complexity}},
\newblock Addison Wesley, 1994.

\bibitem{Bandini2010}
Bandini, S., Manzoni, S., Umeo, H., and Vizzari, G.,
\newblock {\em {Cellullar Automata}},
\newblock Springer-Verlag Berlin Heidelberg, 2010.

\bibitem{Ermentrout93}
Ermentrout, B. and Edelstein-Keshet, L.,
\newblock J. Theor. Biol. {\bf 160} (1993) 97.

\bibitem{cole1996}
Cole, B.~J. and Cheshire, D.,
\newblock Am. Nat. {\bf 148} (1996) 1.

\bibitem{John2004}
John, A., Schadschneider, A., Chowdhury, D., and Nishinari, K.,
\newblock J. Theor. Biol. {\bf 231} (2004) 279.

\bibitem{Altshuler2005}
Altshuler, E., Ramos, O., N\'{u}\~{n}ez, Y., Batista-Leyva, A., and Noda, C.,
\newblock Am. Nat. {\bf 166} (2005) 643.

\bibitem{Noda2006}
Noda, C., Fern\'{a}ndez, J., P\'{e}rez-Penichet, C., and Altshuler, E.,
\newblock Rev. Sci. Inst. {\bf 77} (2006) 126102.

\bibitem{John09}
John, A., Schadschneider, A., Chowdhury, D., and Nishinari, K.,
\newblock Phys. Rev. Lett. {\bf 102} (2009) 108001.

\bibitem{Nicolis2013}
Nicolis, S.~C. et~al.,
\newblock Phys. Rev. Lett. {\bf 110} (2013) 268104.

\bibitem{Jackson2004a}
Jackson, D.~E. and Ratnieks, F. L.~W.,
\newblock Curr. Biol. {\bf 16} (2004) 570.

\bibitem{Farji-brener2010}
Farji-Brener, A.~G., Amador-Vargas, S., Chinchilla, F., Escobar, S., and
  Cabrera, S.,
\newblock Anim. Behav. {\bf 79} (2010) 343.

\bibitem{Prabhakar2012}
Prabhakar, B., Dektar, K.~N., and Gordon, D.~M.,
\newblock PLoS Comp. Biol. {\bf 8} (2012) e1002670.

\bibitem{Halley2001}
Halley, J.~D. and Elgar, M.~A.,
\newblock Aus. J. Zool. {\bf 49} (2001) 59.

\end{thebibliography}

\end{document}